\documentclass[11pt,a4paper,reqno]{amsart}

\usepackage{times}
\usepackage{xspace}
\usepackage[inline]{enumitem}
\usepackage[numbers,sort&compress,longnamesfirst]{natbib}
\usepackage[hmargin=2.0cm,vmargin=2.8cm]{geometry}
\usepackage[colorlinks = true, linkcolor = blue, citecolor = blue]{hyperref}
\usepackage{amsfonts, comment}
\usepackage{amssymb, amsmath, latexsym, mathtools}
\usepackage{amsthm}
\usepackage[capitalize]{cleveref}
\usepackage{color, graphicx} 
\usepackage{todonotes}
\usepackage{lipsum}  
\usepackage{bbm}
\usepackage{algorithm}
\usepackage{algpseudocode}
\usepackage{tabu}
\usepackage{subcaption}

\newlist{myitems}{enumerate}{1}
\setlist[myitems]{label=\arabic*, font=\bfseries, resume}
\numberwithin{equation}{section}
\theoremstyle{plain}

\theoremstyle{definition}

\crefname{assumption}{Assumption}{Assumptions}
\Crefname{assumption}{Assumption}{Assumptions}

\def\<{\langle}
\def\>{\rangle}

\def\ud{\mathrm d}
\DeclareMathOperator*{\argmin}{arg\, min}

\begin{document}

\title[Time Deep Gradient Flow Method for pricing American options]{Time Deep Gradient Flow Method for pricing American options}
\author[J. Rou]{Jasper Rou}

\address{Delft Institute of Applied Mathematics, EEMCS, TU Delft, 2628CD Delft, The Netherlands}
\email{J.G.Rou@tudelft.nl}



\begin{abstract}
In this research, we explore neural network-based methods for pricing multidimensional American put options under the Black–Scholes and Heston models, extending up to five dimensions. We focus on two approaches: the Time Deep Gradient Flow (TDGF) method and the Deep Galerkin Method (DGM). We extend the TDGF method to handle the free-boundary partial differential equation inherent in American options. We carefully design the sampling strategy during training to enhance performance. Both TDGF and DGM achieve high accuracy while significantly outperforming conventional Monte Carlo methods in terms of computational speed. In particular, TDGF tends to be faster during training than DGM.
\end{abstract}

\maketitle

\section{Introduction}
Pricing options is a fundamental problem in financial mathematics. In addition to European options, which can only be exercised at maturity, there exist American options, which can be exercised at any time before maturity. This early exercise feature introduces additional complexity, making the pricing of American options more challenging than European options. One of the first successful methods for pricing American options is the binomial options pricing model introduced by \citet{cox1979option}. Another widely used approach formulates the price of an American option as the solution to a partial differential equation (PDE) with a free boundary or a system of variational inequalities; see \citet{myneni1992pricing} for a comprehensive overview.

As the number of underlying assets increases, the option pricing problem becomes high-dimensional, necessitating more efficient numerical methods. \citet{clarke1999multigrid} describe a multigrid procedure for a fast iterative solution to the pricing of American options. \citet{longstaff2001valuing} proposed a powerful simulation-based technique that approximates the value of American options using least squares regression. \citet{ikonen2008efficient} explored five distinct methods for pricing American options: the projected SOR method, a projected multigrid method, an operator splitting method, a penalty method, and a component-wise splitting method. For an overview of simulation-based methods, see \citet{belomestny2018advanced}.

The development of deep learning has introduced new and powerful ways to solve the problem of pricing American options. Pioneering work in this direction includes \citet{becker2019deep,becker2020pricing,becker2021solving}, who developed deep learning approaches to learn optimal exercise strategies, pricing, and hedging of American options in high-dimensional settings. Other notable contributions include \citet{herrera2021optimal}, who demonstrated the potential of randomized neural networks to outperform traditional deep neural networks and standard basis functions in approximating solutions to optimal stopping problems; \citet{nwankwo2024deep}, who proposed a deep learning framework based on the Landau transformation to handle the free-boundary problem in American option pricing; and \citet{peng2024deep}, who introduced a deep penalty method.

\citet{sirignano2018dgm} proposed the Deep Galerkin Method (DGM), which accurately solves high-dimensional free-boundary PDEs. Recently, \citet{papapantoleon2024time} introduced the Time Deep Gradient Flow (TDGF) method as a more efficient alternative to DGM to solve PDEs arising from European option pricing problems. In this work, we extend the TDGF method to handle free-boundary problems, allowing it to price American options. We compare the performance of DGM and TDGF in pricing American put options under the Black–Scholes and Heston model with up to five underlying assets, evaluating both accuracy and computational efficiency.

The remainder of the paper is organized as follows. \cref{sec:problem}, formulates the problem by defining the system of variational inequalities associated with American options and presenting the multidimensional Black--Scholes and Heston model. \cref{sec:methodology} describes the extension of the TDGF method to American options and introduces the specific neural network architecture and sampling methods used. \cref{sec:numerical} presents numerical results that compare accuracy and computational efficiency. Finally, \cref{sec:conclusion} summarizes our findings.

\section{Problem formulation}
\label{sec:problem}
This section formulates the problem. \cref{sec:american} defines the system of variational inequalities associated with American options. \cref{sec:BS} presents the multidimensional Black--Scholes model. \cref{sec:Heston} presents the multidimensional Heston model.

\subsection{American options}
\label{sec:american}
Let $\mathbf{S} = (S_1, S_2, ..., S_d)$ denote the price processes of $d$ financial assets that evolve according to a diffusion model, and consider an American derivative on $\mathbf{S}$ with payoff $\Psi(\mathbf{S}_t)$ at any time $t<T$, with maturity time $T>0$.
Let $u: [0,T] \times \Omega \to \mathbb{R}$ denote the price of the American derivative, with $\Omega \subseteq \mathbb{R}^d$ and $t$ the time to maturity. Then, $u$ solves the system of inequalities \cite{hilber2013computational}:
\begin{equation}
\label{eq:BS_PDE_american}
\begin{aligned}
    \frac{\partial u}{\partial t} + \mathcal{A} u + r u & \geq 0, \quad & (t, \mathbf{x}) \in [0,T] \times \Omega, \\
    u(t, \mathbf{x}) & \geq \Psi(\mathbf{x}), \quad & (t, \mathbf{x}) \in [0,T] \times \Omega,\\
    u(0, \mathbf{x}) & = \Psi(\mathbf{x}), \quad & \mathbf{x} \in \Omega, \\
    \left( \frac{\partial u}{\partial t} + \mathcal{A} u + r u \right) \left( u(t, \mathbf{x}) - \Psi(\mathbf{x}) \right) & = 0, \quad & (t, \mathbf{x}) \in [0,T] \times \Omega,
\end{aligned}
\end{equation}
with $\mathcal{A}$ a second-order differential operator of the form
\begin{equation}
\label{eq:generator}
    \mathcal{A} u = - \sum_{i,j=1}^d a^{ij} \frac{\partial^2 u}{\partial x_i \partial x_j} + \sum_{i=1}^{d} \beta^i \frac{\partial u}{\partial x_i}.    
\end{equation}
The coefficients $a^{ij}, \beta^i$ of the generator $\mathcal{A}$ relate directly to the dynamics of the stochastic processes $\mathbf{S}$ and can, in general, depend on the time and the spatial variables. 

Problem \eqref{eq:BS_PDE_american} is equivalent to the free-boundary problem:
\begin{equation}
\label{eq:free_boundary}
\begin{aligned}
    \max \left\{ - \frac{\partial u}{\partial t} - \mathcal{A} u - r u,  \Psi(\mathbf{x}) - u(t,\mathbf{x}) \right\} & = 0, \\
    u(0,\mathbf{x}) & = \Psi(\mathbf{x}).
\end{aligned}
\end{equation}

The TDGF reformulates the PDE as an energy minimization problem, which is then approximated in a time-stepping fashion by deep neural networks. In order to write the PDE as an energy minimization problem, we need to split the operator in a symmetric and an (asymmetric) remainder part. Following \citet{papapantoleon2024time}, we can rewrite the operator $\mathcal{A}$ as
\begin{equation}
\label{eq:operator_form}
    \mathcal{A} u = - \nabla \cdot \left( A \nabla u \right) + \mathbf{b} \cdot \nabla u,   
\end{equation}
with a symmetric positive semidefinite matrix
\begin{equation}
\label{eq:coefficients}
A = \begin{bmatrix}
    a^{11} & \dots & a^{d1} \\
    \vdots & \ddots & \vdots \\
    a^{1d} & \dots & a^{dd}
\end{bmatrix}
\quad \textrm{and vector} \quad
\mathbf{b} = \begin{bmatrix}
    b^1 \\
    \vdots \\
    b^d
\end{bmatrix}.
\end{equation}

\subsection{Multidimensional Black--Scholes model}
\label{sec:BS}
In the model by \citet{black1973pricing}, the dynamics of the stock price $S$ follow a geometric Brownian motion. Suppose we have $d$ assets, each following the Black--Scholes model:
\[
\ud S_i(t) = r S_i(t) \ud t + \sigma_i S_i(t) \ud W_i(t)_t, \quad S_i(0) > 0,
\]
with $r>0$ the risk-free rate, $\sigma_i>0$ the volatility of asset $S_i$ and $[W_1(t), ..., W_d(t)]$ Brownian motions with correlation matrix
\[
\begin{bmatrix}
    1 & \rho_{12} & ... & \rho_{1d} \\
    \rho_{12} & 1 & ... & \rho_{2d} \\
    \vdots & \vdots & \ddots & \vdots \\
    \rho_{1d} & \rho_{2d} & ... & 1
\end{bmatrix}.
\]
The generator corresponding to these dynamics, in the form \eqref{eq:generator}, equals
\[
\begin{aligned}
\mathcal{A} u = - \sum_{i=1}^d r S_i \frac{\partial u}{\partial S_i} - \frac{1}{2} \sum_{i=1, j=1}^d \sigma_i \sigma_j S_i S_j \rho_{ij} \frac{\partial^2 u}{\partial S_i \partial S_j}.
\end{aligned}
\]
Applying the product rule gives:
\[
\begin{aligned}
\mathcal{A} u = & - \sum_{i=1}^d r S_i \frac{\partial u}{\partial S_i}  - \frac{1}{2} \sum_{i=1}^d \sigma_i^2 S_i^2 \frac{\partial^2 u}{\partial S_i^2} - \frac{1}{2} \sum_{i=1}^d \sum_{j \neq i} \sigma_i \sigma_j S_i S_j \rho_{ij} \frac{\partial^2 u}{\partial S_i \partial S_j} \\
= & - \sum_{i=1}^d r S_i \frac{\partial u}{\partial S_i} - \frac{1}{2} \sum_{i=1}^d \frac{\partial}{\partial S_i} \left( \sigma_i^2 S_i^2 \frac{\partial u}{\partial S_i} \right) + \sum_{i=1}^d \sigma_i^2 S_i \frac{\partial u}{\partial S_i} - \frac{1}{2} \sum_{i=1}^d \sum_{j \neq i} \frac{\partial}{\partial S_j} \left( \sigma_i \sigma_j S_i S_j \rho_{ij} \frac{\partial u}{\partial S_i} \right) \\
& + \frac{1}{2} \sum_{i=1}^d \sum_{j \neq i} \sigma_i \sigma_j S_i \rho_{ij} \frac{\partial u}{\partial S_i} \\
= & \sum_{i=1}^d \left( \sigma_i^2 + \frac{1}{2} \sum_{j \neq i} \sigma_i \sigma_j \rho_{ij} - r \right) S_i \frac{\partial u}{\partial S_i} - \frac{1}{2} \sum_{i,j=1}^d \frac{\partial}{\partial S_j} \left( \sigma_i \sigma_j S_i S_j \rho_{ij} \frac{\partial u}{\partial S_i} \right).
\end{aligned}
\]
Therefore, the operator $\mathcal{A}$ takes the form \eqref{eq:operator_form} with the coefficients in \eqref{eq:coefficients} provided by 
\begin{align*}
    a^i &= \frac{1}{2} \sigma_i \sigma_j S_i S_j \rho_{ij}, \quad & i=1, ..., d, \\
    b^i &= \left( \sigma_i^2 + \frac{1}{2} \sum_{j \neq i} \sigma_i \sigma_j \rho_{ij} - r \right) S_i, \quad & i=1, ..., d.
\end{align*}

\subsection{Multidimensional Heston model}
\label{sec:Heston}
The model by \citet{heston1993closed} is a popular stochastic volatility model. In $d$ dimensions the dynamics of asset $S$ and variance process $V$ are
\[
\begin{aligned}
\ud S_i(t) &= r S_i(t) \ud t + \sqrt{V_i(t)} S_i(t) \ud W_i(t), \quad && S_i(0) > 0, \\
\ud V_i(t) &= \lambda_i \left( \kappa_i - V_i(t) \right) \ud t + \eta_i \sqrt{V_i(t)} \ud B_i(t), \quad && V_i(0) > 0.
\end{aligned}
\]
Here and $\lambda, \kappa, \eta \in \mathbb{R}_+$ and $[B_1(t), ..., B_d(t), W_1(t), ..., W_d(t)]$ are Brownian motions, with correlation matrix \cite{wadman2010advanced}:
\[
\Sigma = \begin{bmatrix}
    I_d & \Sigma_{SV} \\
    \Sigma_{SV}^T & \Sigma_S
\end{bmatrix} ,
\Sigma_{SV} = \begin{bmatrix}
    \rho_1 & 0 & ... & 0 \\
    0 & \rho_2 & ... & 0 \\
    \vdots & \vdots & \ddots & \vdots \\
    0 & 0 & ... & \rho_d
\end{bmatrix},
\Sigma_{S} = \begin{bmatrix}
    \rho_{11} & \rho_{12} & ... & \rho_{1d} \\
    \rho_{21} & \rho_{22} & ... & \rho_{2d} \\
    \vdots & \vdots & \ddots & \vdots \\
    \rho_{d1} & \rho_{d2} & ... & \rho_{dd},
\end{bmatrix}
\]
with $\rho_i$ the correlation between $W_i$ and $B_i$ and $\rho_{ij}$ the correlation between $W_i$ and $W_j$. The correlations between the $B_i$ and between $W_j$ and $B_i$ are 0.

Let $f \left( S_1(t),...,S_d(t),V_1(t),...,V_d(t) \right): \mathbb{R}^{2d} \to \mathbb{R}$ be a $\mathcal{C}^2$-function. Then Itô's formula gives
\[
\begin{aligned}
& \ud f \left( S_1(t),...,S_d(t),V_1(t),...,V_d(t) \right) \\
= & \sum_{i=1}^d \frac{\partial f}{\partial S_i} \ud S_i + \sum_{i=1}^d \frac{\partial f}{\partial V_i} \ud V_i + \frac{1}{2} \sum_{i,j=1}^d \frac{\partial^2 f}{\partial S_i \partial S_j} \ud \left \langle S_i, S_j \right \rangle + \sum_{i,j=1}^d \frac{\partial^2 f}{\partial S_i \partial V_j} \ud \left \langle S_i, V_j \right \rangle  \\
& + \frac{1}{2} \sum_{i,j=1}^d \frac{\partial^2 f}{\partial V_i \partial V_j} \ud \left \langle V_i, V_j \right \rangle \\
= & \sum_{i=1}^d \frac{\partial f}{\partial S_i} r S_i \ud t 
+ \sum_{i=1}^d \frac{\partial f}{\partial V_i} \lambda_i \left( \kappa_i - V_i \right) \ud t 
+ \frac{1}{2} \sum_{i,j=1}^d \frac{\partial^2 f}{\partial S_i \partial S_j} \rho_{ij} \sqrt{V_i V_j} S_i S_j \ud t
+ \sum_{i=1}^d \frac{\partial^2 f}{\partial S_i \partial V_i} V_i S_i \eta_i \rho_i \ud t \\
& + \frac{1}{2} \sum_{i=1}^d \frac{\partial^2 f}{\partial V_i^2} \eta_i^2 V_i \ud t
+ \textrm{martingale}.
\end{aligned}
\]
Then the generator corresponding to these dynamics, in the form \eqref{eq:generator}, equals
\[
\begin{aligned}
    \mathcal{A} u = & - \sum_{i=1}^d \frac{\partial u}{\partial S_i} r S_i - \sum_{i=1}^d \frac{\partial u}{\partial V_i} \lambda_i \left( \kappa_i - V_i \right) - \frac{1}{2} \sum_{i,j=1}^d \frac{\partial^2 u}{\partial S_i \partial S_j} \rho_{ij} \sqrt{V_i V_j} S_i S_j - \sum_{i=1}^d \frac{\partial^2 u}{\partial S_i \partial V_i} V_i S_i \eta_i \rho_i \\
    & - \frac{1}{2} \sum_{i=1}^d \frac{\partial^2 u}{\partial V_i^2} \eta_i^2 V_i.
\end{aligned}
\]
Applying the product rule gives:
\[
\begin{aligned}
 \mathcal{A} u = & 
 -\sum_{i=1}^d \frac{\partial u}{\partial S_i} r S_i 
 - \sum_{i=1}^d \frac{\partial u}{\partial V_i} \lambda_i \left( \kappa_i - V_i \right) 
 - \frac{1}{2} \sum_{i=1}^d \frac{\partial^2 u}{\partial S_i^2} V_i S_i^2 
 - \frac{1}{2} \sum_{i=1}^d \sum_{j \neq i}^d \frac{\partial^2 u}{\partial S_i \partial S_j} \rho_{ij} \sqrt{V_i V_j} S_i S_j \\
 & - \sum_{i=1}^d \frac{\partial^2 u}{\partial S_i \partial V_i} V_i S_i \eta_i \rho_i 
 - \frac{1}{2} \sum_{i=1}^d \frac{\partial^2 u}{\partial V_i^2} \eta_i^2 V_i \\
 = & - \sum_{i=1}^d \frac{\partial u}{\partial S_i} r S_i 
 - \sum_{i=1}^d \frac{\partial u}{\partial V_i} \lambda_i \left( \kappa_i - V_i \right) 
 - \frac{1}{2} \sum_{i=1}^d \frac{\partial}{\partial S_i} \left( \frac{\partial u}{\partial S_i} V_i S_i^2 \right) 
 + \sum_{i=1}^d \frac{\partial u}{\partial S_i} V_i S_i \\
 & - \frac{1}{2} \sum_{i=1}^d \sum_{j \neq i}^d \frac{\partial}{\partial S_j} \left( \frac{\partial u}{\partial S_i} \rho_{ij} \sqrt{V_i V_j} S_i S_j \right)
 + \frac{1}{2} \sum_{i=1}^d \sum_{j \neq i}^d \frac{\partial u}{\partial S_i} \rho_{ij} \sqrt{V_i V_j} S_i 
 - \frac{1}{2} \sum_{i=1}^d \frac{\partial}{\partial S_i} \left( \frac{\partial u}{\partial V_i} V_i S_i \eta_i \rho_i \right) \\
 & + \frac{1}{2} \sum_{i=1}^d \frac{\partial u}{\partial V_i} V_i \eta_i \rho_i
 - \frac{1}{2} \sum_{i=1}^d \frac{\partial}{\partial V_i} \left( \frac{\partial u}{\partial S_i} V_i S_i \eta_i \rho_i \right) - \frac{1}{2} \sum_{i=1}^d \frac{\partial u}{\partial S_i} S_i \eta_i \rho_i
 - \frac{1}{2} \sum_{i=1}^d \frac{\partial}{\partial V_i} \left( \frac{\partial u}{\partial V_i} \eta_i^2 V_i \right) \\
 & + \frac{1}{2} \sum_{i=1}^d \frac{\partial u}{\partial V_i} \eta_i^2  \\
  = & \sum_{i=1}^d \left( \frac{1}{2} \left( V_i + \sum_{j=1}^d \rho_{ij} \sqrt{V_i V_j} + \eta_i \rho_i \right) -r \right) S_i \frac{\partial u}{\partial S_i}
 + \sum_{i=1}^d \left( \lambda_i \left( V_i - \kappa_i \right) + \frac{1}{2} V_i \eta_i \rho_i + \frac{1}{2} \eta_i^2 \right) \frac{\partial u}{\partial V_i} \\
 & - \frac{1}{2} \sum_{i,j=1}^d \frac{\partial}{\partial S_j} \left( \frac{\partial u}{\partial S_i} \rho_{ij} \sqrt{V_i V_j} S_i S_j \right)
 - \frac{1}{2} \sum_{i=1}^d \frac{\partial}{\partial S_i} \left( \frac{\partial u}{\partial V_i} V_i S_i \eta_i \rho_i \right) 
 - \frac{1}{2} \sum_{i=1}^d \frac{\partial}{\partial V_i} \left( \frac{\partial u}{\partial S_i} V_i S_i \eta_i \rho_i \right) \\
 & - \frac{1}{2} \sum_{i=1}^d \frac{\partial}{\partial V_i} \left( \frac{\partial u}{\partial V_i} \eta_i^2 V_i \right). \\
\end{aligned}
\]
Therefore, the operator $\mathcal{A}$ takes the form \eqref{eq:operator_form} with the coefficients in \eqref{eq:coefficients} provided by 
\begin{align*}
    a^{ij} &= \frac{1}{2} \rho_{ij} \sqrt{V_i V_j} S_i S_j, \quad & i,j=1,...,d, \\
    a^{ji} = a^{ij} &= \frac{1}{2} V_i S_i \eta_i \rho_i, \quad & i=1, ..., d, j=i+d, \\
    a^{ii} &= \frac{1}{2} \eta_i^2 V_i, \quad & i=d+1,...,2d, \\
    a^{ij} &= 0, \quad & \text{otherwise}, \\
    b^i &= \left( \frac{1}{2} \left( V_i + \sum_{j=1}^d \rho_{ij} \sqrt{V_i V_j} + \eta_i \rho_i \right) -r \right) S_i, \quad & i=1,...,d, \\
    b^i &= \lambda_i \left( V_i - \kappa_i \right) + \frac{1}{2} V_i \eta_i \rho_i + \frac{1}{2} \eta_i^2, \quad & i=d+1,...,2d.
\end{align*}

\section{Methodology}
\label{sec:methodology}
This section provides the details on how to solve the problem from the previous section. \cref{sec:time} describes the extension of the TDGF method to American options. \cref{sec:architecture} introduces the specific neural network architecture used. \cref{sec:sampling} introduces the specific sampling methods used.

\subsection{Time Deep Gradient Flow Method}
\label{sec:time}
The TDGF is a neural network method to efficiently solve high-dimensional PDEs \cite{papapantoleon2024time, georgoulis2023discrete}. 
Let us divide the time interval $[0,T]$ into $K$ equally spaced intervals $(t_{k-1},t_k]$, with $h = t_k - t_{k-1} = \frac{1}{K}$ for $k=0,1,\dots,K$. 
By first discretizing the PDE in time and then rewriting the discretized PDE as an energy functional we can approximate the solution to the PDE
\[
\begin{aligned}
\frac{\partial u}{\partial t} - \nabla \cdot \left( A \nabla u \right) + \mathbf{b} \cdot \nabla u + r u & = 0, \\
u(0) & = \Psi,
\end{aligned}
\]
by
\[
\begin{aligned}
u(t_k, \mathbf{x}) \approx & U^k = \argmin I^k (u), \\
I^k(u) = & \frac{1}{2} \left \Vert u - U^{k-1} \right \Vert_{L^2(\Omega)}^2 + h \left( \int_{\Omega} \frac{1}{2} \left( \left( \nabla u \right)^\mathsf{T} A \nabla u + r u^2 \right) + F \left( U^{k-1} \right) u \ud x \right), \\
U^0 =& \Psi.
\end{aligned}
\]
Let $f^k(\mathbf{x}; \theta)$ denote a neural network approximation of $U^k$ with trainable parameters $\theta$. 
Applying a Monte Carlo approximation to the integrals, the discretized cost functional takes the form
\begin{align*}
I^k\left(f^k(\mathbf{x};\theta)\right) \approx L^k \left( \theta ; \mathbf{x} \right) 
    & = \frac{|\Omega|}{2M} \sum_{m=1}^{M} \left( f^k(\mathbf{x}_m;\theta) - f^{k-1}(\mathbf{x}_m) \right)^2 + h N^k \left( \theta ; \mathbf{x} \right),
\end{align*}
with
\begin{align*}
N^k \left( \theta ; \mathbf{x} \right) 
    = & \frac{|\Omega|}{M} \sum_{m=1}^{M} \Bigg[ \frac{1}{2} \left( \left( \nabla f^k(\mathbf{x}_m; \theta) \right)^\mathsf{T} A \nabla f^k(\mathbf{x}_m; \theta) + r \left( f^k({\mathbf{x}_m}; \theta) \right) ^2 \right) \\
    & + \left( \mathbf{b} \cdot \nabla f^{k-1}(\mathbf{x}_m) \right) f^k(\mathbf{x}_m; \theta) \Bigg].
\end{align*}
Here, $M$ denotes the number of samples $\mathbf{x}_m$.
From equation \eqref{eq:BS_PDE_american}, the PDE is satisfied if the solution $u$ is strictly larger than the payoff $\Psi$.
Therefore, we only train the PDE on the part of the domain where the solution is above the payoff.

In order to minimize this cost function, we use a stochastic gradient descent type algorithm, \textit{i.e.} an iterative scheme of the form: 
\[
    \theta_{n+1} = \theta_n - \alpha_n \nabla_{\theta} L^k(\theta_n; \mathbf{x}).
\] 
The hyperparameter $\alpha_n$ is the step size of our update, called the learning rate. 
An overview of the TDGF appears in Algorithm \ref{alg:time_DNM}.

\begin{algorithm}
\caption{Time Deep Gradient Flow method for American Options}\label{alg:time_DNM}
\begin{algorithmic}[1]
\State Initialize $\theta_0^0$.
\For{each sampling stage $n$}
\State Generate random points $\mathbf{x}_m$ for training.
\State Calculate the cost functional $L^0(\theta_n^0; \mathbf{x}) = \frac{1}{M} \sum_{m=1}^M \left( f^0(\mathbf{x}_m; \theta^0_n) - \Psi(\mathbf{x}_m) \right)^2$ for the selected points.
\State Take a descent step $\theta_{n+1}^0 = \theta_n^0 - \alpha_n \nabla_{\theta} L^0(\theta_n^0; \mathbf{x})$.
\EndFor
\For{each time step $k = 1,\dots,K$}
\State Initialize $\theta_0^k = \theta^{k-1}$.
\For{each sampling stage $n$}
\State Generate random points $\mathbf{x}_m$ for training.
\State Select the points $\mathbf{x}_m$ where $f^k(\mathbf{x}_m) > \Psi(\mathbf{x}_m)$.
\State Calculate the cost functional $L^k(\theta_n^k; \mathbf{x})$ for the selected points.
\State Take a descent step $\theta_{n+1}^k = \theta_n^k - \alpha_n \nabla_{\theta} L^k(\theta_n^k; \mathbf{x})$.
\EndFor
\EndFor
\end{algorithmic}
\end{algorithm}

\subsection{Architecture}
\label{sec:architecture}
Let us now describe some details about the design of the neural network architecture and the implementation of the numerical method.
We would like to use information about the option price in order to facilitate the training of the neural network. 
The price of an American option can be decomposed in two (positive) values: the intrinsic value and the continuation value. The intrinsic value is the value of the option if we exercise, which we know to be $\Psi$. The continuation value is the value of the option if we do not exercise and let the stock continue following the PDE. 
From the second line of \eqref{eq:BS_PDE_american} we know $u \geq \Psi$. The neural network learns the continuation value, instead of the option price itself. 

The architecture of the neural network for the TDGF method follows that of the DGM \cite{sirignano2018dgm}. Overall, we set:
\[
\begin{aligned}
    X^1 & = \sigma_1 \left( W^1 \mathbf{x} + b^1 \right), & \\
    Z^{l+1} & = \sigma_1 \left( U^{z,l} \mathbf{x} + W^{z,l} X^l + b^{z,l} \right), \quad & l=1,\dots,L, \\
    G^{l+1} & = \sigma_1 \left( U^{g,l} \mathbf{x} + W^{g,l} X^l + b^{g,l} \right), \quad & l=1,\dots,L, \\
    R^{l+1} & = \sigma_1 \left( U^{r,l} \mathbf{x} + W^{r,l} X^l + b^{r,l} \right), \quad & l=1,\dots,L, \\
    H^{l+1} & = \sigma_1 \left( U^{h,l} \mathbf{x} + W^{h,l} \left( X^l \odot R^l \right) + b^{h,l} \right), \quad & l=1,\dots,L, \\
     X^{l+1} & = \left( 1 - G^l \right) \odot H^l + Z^l \odot X^{l}, \quad & l=1,\dots,L, \\
    f(\mathbf{x}; \theta) & = \Psi + \sigma_2 \left( W X^{L+1} + b \right).  &
\end{aligned}
\]
Here, $L$ is the number of hidden layers, $\sigma_i$ is the activation function for $i=1,2$, and $\odot$ denotes the element-wise multiplication. 
In the numerical experiments, we use 3 layers and 50 neurons per layer. 
The activation functions are the hyperbolic tangent function, $\sigma_1(x) = \tanh(x)$, and the softplus function, $\sigma_2(x) = \log \left( e^x +1 \right)$, which guarantees that the option price remains above the no-arbitrage bound. 

We consider a maturity of $T=1.0$ year, and take the number of time steps equal to $K = 100$. 
We use $2000$ sampling stages in each time step.
For the optimization we use Adam algorithm \cite{kingma2014adam} with a learning rate $\alpha = 3 \times 10^{-4}$, $(\beta_1,\beta_2) = (0.9,0.999)$ and zero weight decay.
The training is performed on the DelftBlue supercomputer \cite{DHPC2024}, using a single NVidia Tesla V100S GPU.

\subsection{Sampling}
\label{sec:sampling}
\begin{figure}
    \centering
    \begin{subfigure}{0.49\textwidth}
        \centering
        \includegraphics[width=\linewidth]{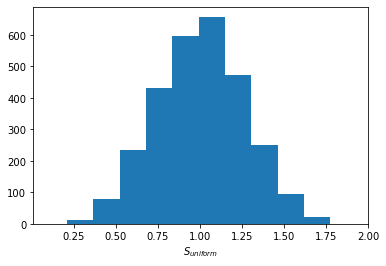}
        \caption{Uniform}
        \label{fig:sampling_d=5_uniform}
    \end{subfigure}
    \begin{subfigure}{0.49\textwidth}
        \centering
        \includegraphics[width=\linewidth]{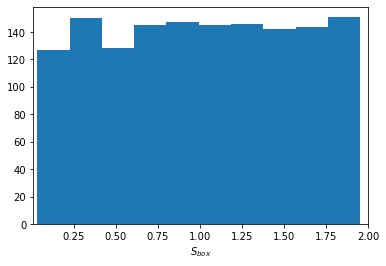}
        \caption{19 boxes}
        \label{fig:sampling_d=5_box}
    \end{subfigure}
    \caption{Histogram of the moneyness for 5 dimensions with 2850 samples for different sampling methods.}
    \label{fig:sampling_d=5}
\end{figure}

For the sampling we have to be particularly careful in the multidimensional case. We assign equal weight to each asset, therefore the moneyness of the option is the average of the individual moneynesses:
\[
S = \frac{1}{d} \sum_{i=1}^d S_i.
\]
If we sample each $S_i$ uniformly we obtain a histogram of the moneyness as in Figure \ref{fig:sampling_d=5_uniform}. There are barely samples at the edges of the domain, therefore the network does not learn the solution in this area.

\begin{figure}
    \centering
    \includegraphics[width=\linewidth]{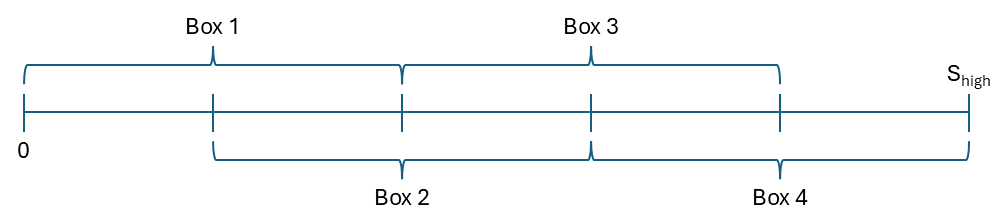}
    \caption{Sampling domain with 4 boxes.}
    \label{fig:box_sampling}
\end{figure}

To cope with this issue, we split the domain in $n-1$ smaller boxes \[
\left[ 0, \frac{2 S_{high}}{n} \right], \left[ \frac{S_{high}}{n}, \frac{3 S_{high}}{n} \right], ..., \left[ \frac{(n-2) S_{high}}{n}, S_{high} \right]
\]
and take samples from each box separately. Figure \ref{fig:box_sampling} displays an example of the domain and boxes for $n=5$. Figure \ref{fig:sampling_d=5_box} displays the moneyness using this box sampling.

For the TDGF, during training of the time steps we are only concerned with points where the neural network is larger than the payoff. Therefore, for the TDGF we apply initial training with box sampling and during the time steps we apply uniform sampling.
In each sampling stage we take $30$ samples per box per dimension ($30d$ for Black--Scholes, $60d$ for Heston) and use 19 boxes. 

In the experiments, we choose the parameters such that in the Black--Scholes model, the continuation value is larger than in the Heston model. In the Heston model, the continuation value is already zero for moneyness larger than 1.5, while for Black--Scholes, the continuation value can be positive for moneyness beyond 2. Numerical experiments suggest therefore that for better results we consider the sampling domain of the moneyness $S \in [0.01,3.0]$ for Black--Scholes and $S \in [0.01,2.0]$ for Heston.
The domain of the Heston volatility is $V \in [0.001, 0.1]$.

\section{Numerical results}
\label{sec:numerical}
Since we do not consider dividends and assume $r \geq 0$ the best exercise strategy for an American call option is to wait until maturity. Therefore, the price of an American call is the same as a European call \cite{musiela2005american}. So we consider an American basket put with equal weights for each asset:
$\Psi(\mathbf{S}) = \left( K - \frac{1}{d} \sum_{i=1}^d S_i \right)^+.$

We compare the TDGF method with the DGM \cite{sirignano2018dgm}. 
In the DGM approach, In the DGM approach, we minimize the $L^2$-error of the free-boundary PDE:
\[
\begin{aligned}
\left \Vert \max \left\{ - \frac{\partial u}{\partial t} - \mathcal{A} u - r u,  \Psi(\mathbf{x}) - u(t, \mathbf{x}) \right\} \right \Vert_{L^2([0,T] \times \Omega)}^2 
+ \left \Vert u(0,\mathbf{x}) - \Psi(\mathbf{x}) \right \Vert_{L^2(\Omega)}^2. 
\end{aligned}
\]
To have a fair comparison between the two methods, we use 200,000 sampling stages and the same learning rate $\alpha = 3 \times 10^{-4}$ for the DGM. 

In order to evaluate the accuracy of the two methods, we need a reference value. 
We compute 1,000 Monte Carlo paths with 1,000 time steps and apply the method of \citet{longstaff2001valuing}, which applies a polynomial regression of order 4 on the paths where the intrinsic value is positive.

When evaluating, we plot the continuation value against the moneyness on an equidistant grid of 47 points where the moneyness and volatility in each dimension are the same.

\subsection{Accuracy}
\begin{figure}
    \centering
    \includegraphics[width=0.95\linewidth]{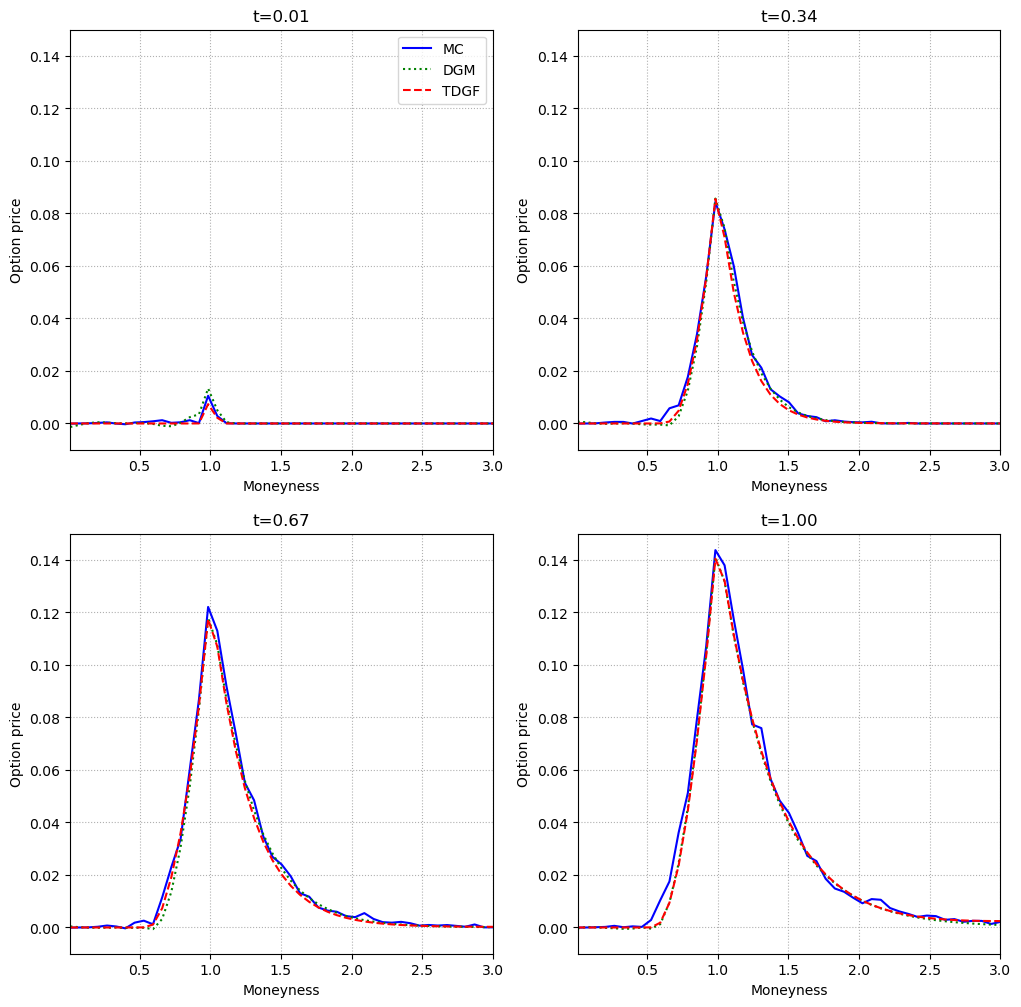}
    \caption{Difference between the option price and the payoff in the two-dimensional Black--Scholes model against the moneyness of the stock, compared to the DGM and Monte Carlo with Longstaff--Schwartz methods, for four different times to maturity with $r=0.05$ and $\sigma_i = 0.5$ and $\rho_{ij}=0.5$ for each $i$ and $j$.}
    \label{fig:BS_d=2}
\end{figure}

\begin{figure}
    \centering
    \includegraphics[width=0.95\linewidth]{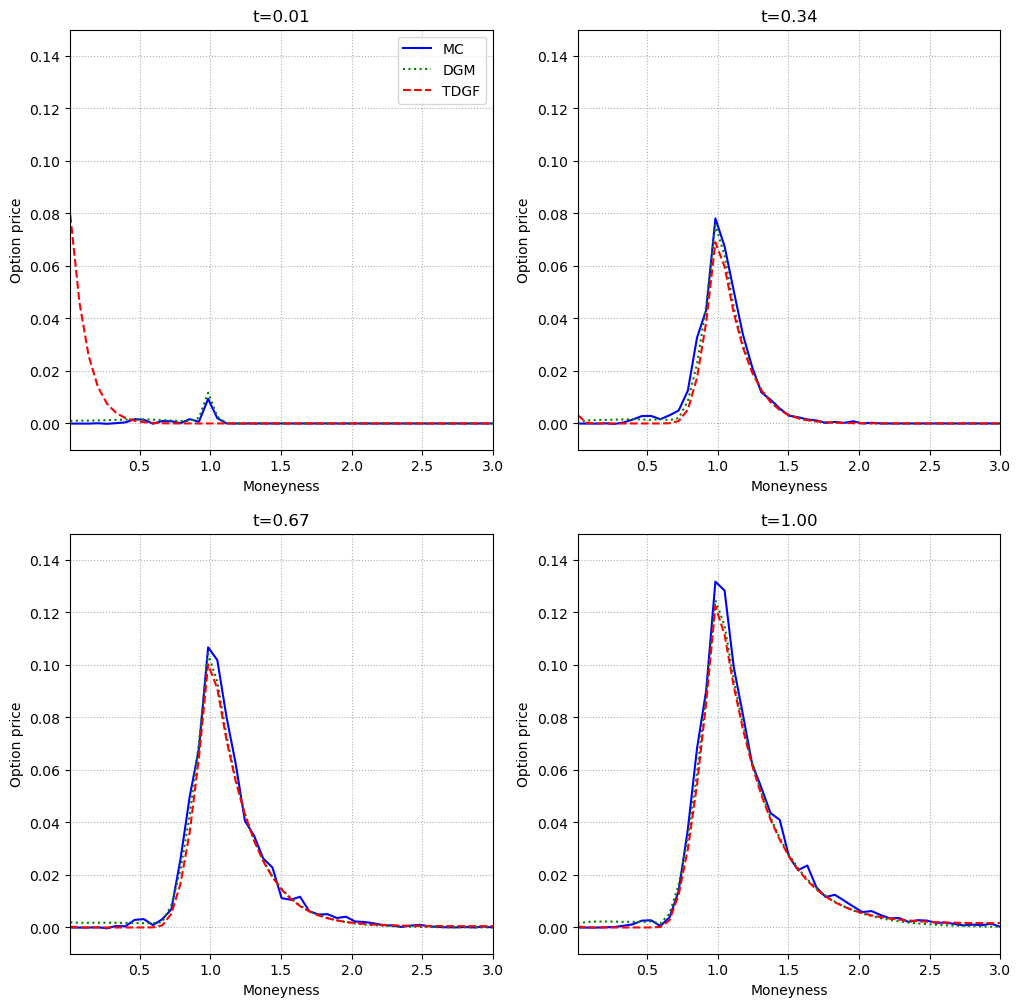}
    \caption{Difference between the option price and the payoff in the five-dimensional Black--Scholes model against the moneyness of the stock, compared to the DGM and Monte Carlo with Longstaff--Schwartz methods, for four different times to maturity with $r=0.05$ and $\sigma_i = 0.5$ and $\rho_{ij}=0.5$ for each $i$ and $j$.}
    \label{fig:BS_d=5}
\end{figure}

Figure \ref{fig:BS_d=2} presents the difference between the option price and the payoff against moneyness in the two-dimensional Black--Scholes model.
Figure \ref{fig:BS_d=5} presents the difference between the option price and the payoff against moneyness in the five-dimensional Black--Scholes model. All three methods display similar values and therefore both DGM and TDGF give accurate results.

\begin{figure}
    \centering
    \includegraphics[width=0.95\linewidth]{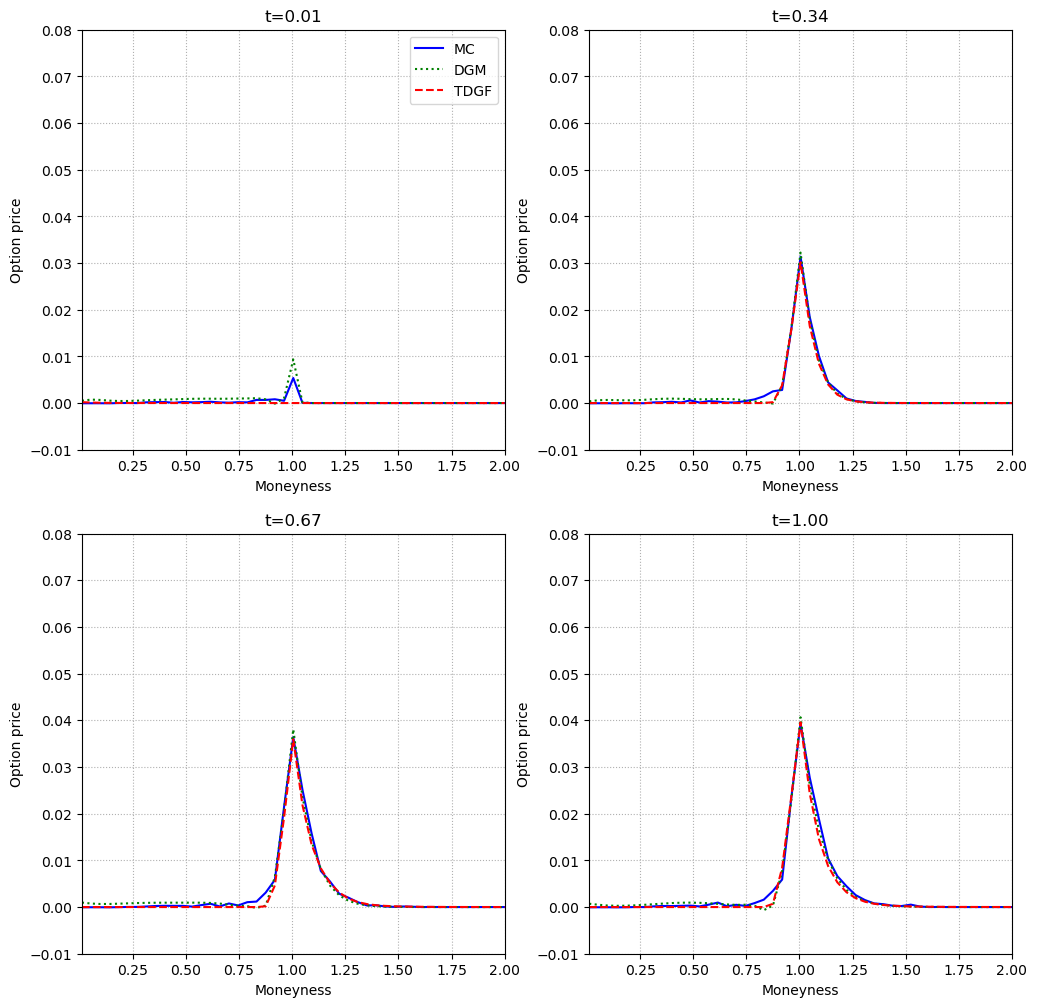}
    \caption{Difference between the option price and the payoff in the two-dimensional Heston model against the moneyness of the stock, compared to the DGM and Monte Carlo with Longstaff--Schwartz methods, for four different times to maturity with $r=0.05$ and $\eta_i = 0.1$, $\rho_{ij}=0.5$, $\rho_i=-0.5$, $\kappa_i=0.01$, $V^i_0 = 0.05$ and $\lambda_i = 2.0$ for each $i$ and $j$.}
    \label{fig:Heston_d=2}
\end{figure}

\begin{figure}
    \centering
    \includegraphics[width=0.95\linewidth]{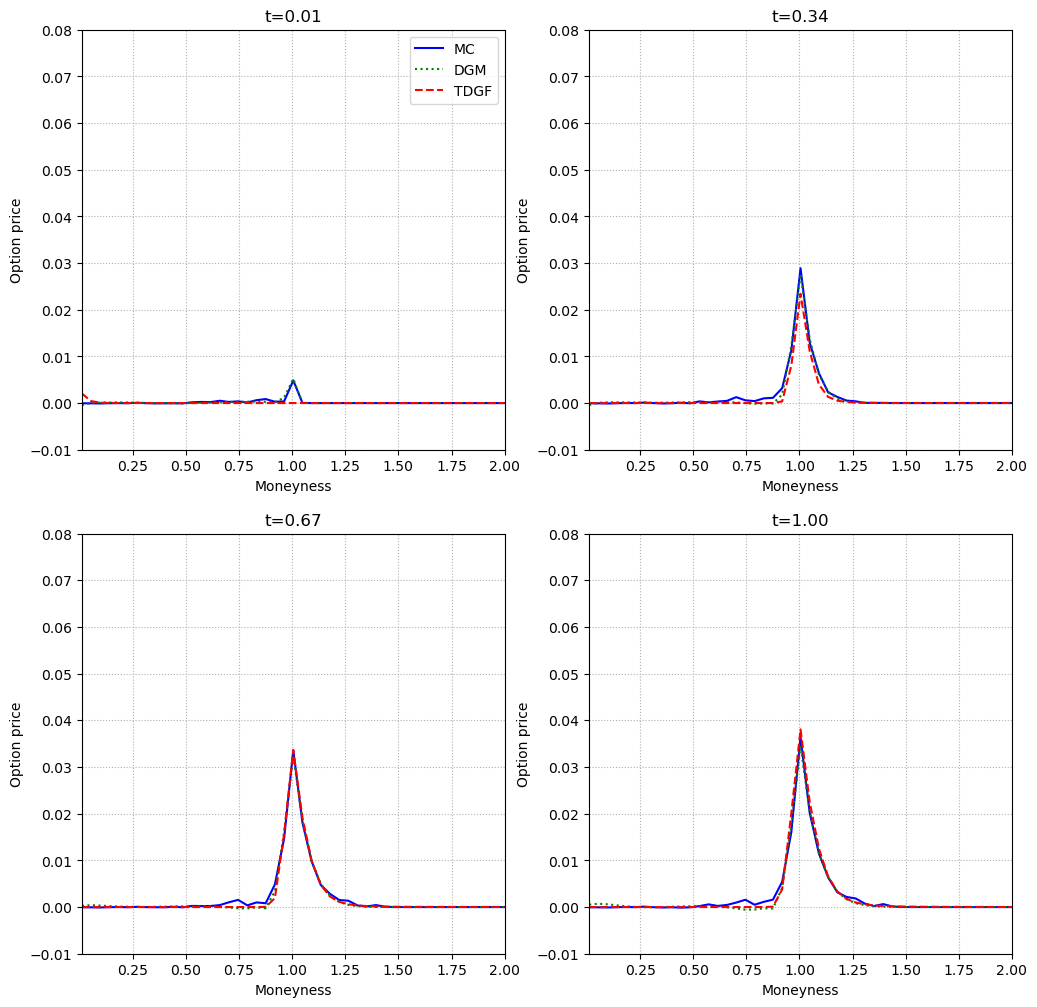}
    \caption{Difference between the option price and the payoff in the five-dimensional Heston model against the moneyness of the stock, compared to the DGM and Monte Carlo with Longstaff--Schwartz methods, for four different times to maturity with $r=0.05$ and $\eta_i = 0.1$, $\rho_{ij}=0.5$, $\rho_i=-0.5$, $\kappa_i=0.01$, $V^i_0 = 0.05$ and $\lambda_i = 2.0$ for each $i$ and $j$.}
    \label{fig:Heston_d=5}
\end{figure}

Figure \ref{fig:Heston_d=2} presents the difference between the option price and the payoff against moneyness in the two-dimensional Heston model.
Figure \ref{fig:Heston_d=5} presents the difference between the option price and the payoff against moneyness in the five-dimensional Heston model. All three methods display similar values and therefore both DGM and TDGF give accurate results. 

\subsection{Running times}
Table \ref{tab:training_time} summarizes the training times for the TDGF and the DGM methods in the different models. 
As expected, due to the time stepping and the absence of a second derivative in the cost function, the training of the TDGF method is faster than for the DGM method.

\begin{table}
    \centering
    \begin{tabu}{l|r|r|r|r} \hline
        Model & Black--Scholes $d=2$ & Black--Scholes $d=5$ & Heston $d=2$ & Heston $d=5$ \\ \hline
        DGM & 8293 & 16174 & 17997 & 41718 \\ \hline
        TDGF & 4543 & 6583 & 7138 & 12881 \\ \hline
    \end{tabu}
    \caption{Training time in seconds of the different methods for an American put option in the different models.}
    \label{tab:training_time}
\end{table}

Table \ref{tab:computing_time} presents the computational times for the TDGF and the DGM in all models. 
The computational times are the average over 34 computations at different time points. 
Both methods are significantly faster than the Monte Carlo method. 
\begin{table}
    \centering
    \begin{tabu}{l|l|l|l|l} \hline
        Model & Black--Scholes $d=2$ & Black--Scholes $d=5$ & Heston $d=2$ & Heston $d=5$ \\ \hline
        MC & 4.6 & 4.5 & 6.0 & 6.6 \\ \hline
        DGM & 0.0015 & 0.0024 & 0.0015 & 0.0016 \\ \hline
        TDGF & 0.0018 & 0.0017 & 0.0016 & 0.0017 \\ \hline
    \end{tabu}
    \caption{Computational time in seconds of the different methods for an American put option in the different models.}
    \label{tab:computing_time}
\end{table}

\section{Conclusion}
\label{sec:conclusion}
In this research, we explored neural network-based methods for pricing multidimensional American put options under the Black–Scholes and Heston models, extending up to five dimensions. We focused on two approaches: the TDGF method and the DGM.

We extended the TDGF method to handle the free-boundary PDE inherent in American options. We restricted training to the region where the PDE holds and incorporated the lower bound constraint directly into the network architecture. Additionally, we carefully designed the sampling strategy during training to enhance performance.

Both TDGF and DGM achieve high accuracy while significantly outperforming conventional Monte Carlo methods in terms of computational speed. Notably, TDGF tends to be faster during training than DGM. 

\bibliographystyle{abbrvnat}
\bibliography{references}

\end{document}